\documentclass[pdftex,twocolumn,epjc3]{svjour3}          

\RequirePackage[T1]{fontenc}

\smartqed  

\RequirePackage{graphicx}
\RequirePackage{amssymb}
\RequirePackage{bm}
\RequirePackage{mathptmx}      
\RequirePackage{flushend}
\RequirePackage{cuted}
\RequirePackage[numbers,sort&compress]{natbib}
\RequirePackage[colorlinks,citecolor=blue,urlcolor=blue,linkcolor=blue]{hyperref}

\journalname{Eur. Phys. J. C}

\newcommand{\be}[1]{\begin{equation}\label{#1}}
\newcommand{\ee}{\end{equation}}
\newcommand{\ba}[1]{\begin{eqnarray}\label{#1}}
\newcommand{\ea}{\end{eqnarray}}
\newcommand{\rf}[1]{(\ref{#1})}
\newcommand{\nn}{\nonumber}

\newcommand{\for}{\quad\mbox{\rm for}\quad}

\newcommand{\const}{\mbox{\rm const}\,}

\if@twocolumn
    \newenvironment{widetext}
        {%
            \begin{strip}
            \rule{\dimexpr(0.5\textwidth-0.5\columnsep-0.4pt)}{0.4pt}%
            \rule{0.4pt}{6pt}
            \par 
            \parindent \@parindent
        }%
        {%
            \par
            \hfill\rule[-6pt]{0.4pt}{6.4pt}%
            \rule{\dimexpr(0.5\textwidth-0.5\columnsep-1pt)}{0.4pt}
            \end{strip}
        }
\else
    
\fi

\begin{document}

\title{Effects of nonlinearity of $f(R)$ gravity and perfect fluid in Kaluza-Klein models with spherical compactification}


\author{Ezgi Canay\thanksref{e1,addr1}
        \and
        Maxim Eingorn\thanksref{e2,addr2} 
        \and
        Alexander Zhuk\thanksref{e3,addr3}
}

\thankstext{e1}{e-mail:ezgicanay@itu.edu.tr}
\thankstext{e2}{e-mail:maxim.eingorn@gmail.com}
\thankstext{e3}{e-mail:ai.zhuk2@gmail.com}

\institute{Department of Physics, Istanbul Technical University, Maslak 34469 Istanbul, Turkey\label{addr1}
          \and
          Department of Mathematics and Physics, North Carolina Central University, Fayetteville st. 1801, Durham, North Carolina 27707, U.S.A.\label{addr2}
          \and
          Astronomical Observatory, Odessa National University, Dvoryanskaya st. 2, Odessa 65082, Ukraine\label{addr3}
}

\date{Received: date / Accepted: date}

\maketitle

\begin{abstract}
We study the effects associated with nonlinearity of $f(R)$ gravity and of the background perfect fluid manifested in the Kaluza-Klein model with spherical compactification. The background space-time is perturbed by a massive gravitating source which is pressureless in the external space but has an arbitrary equation of state (EoS) parameter in the internal space. As characteristics of a nonlinear perfect fluid, the squared speeds of sound are not equal to the background EoS parameters in the external and internal spaces. In this setting, we find exact solutions to the linearized Einstein equations for the perturbed metric coefficients. For nonlinear models with $f^{\prime\prime}(R_0)\neq0$, we show that these coefficients acquire correction terms in the form of two summed Yukawa potentials and that in the degenerated case, the solutions are reduced to a single Yukawa potential with some ``corrupted'' prefactor (in front of the exponential function), which, in addition to the standard $1/r$ term, contains a contribution independent of the  three-dimensional distance $r$. In the linear $f''(R)=0$ model, we generalize the previous studies to the case of an arbitrary nonlinear perfect fluid. We also investigate the particular case of the nonlinear background perfect fluid with zero speed of sound in the external space and demonstrate that a non-trivial solution exists only in the case of $f''(R_0)=0$. 
\end{abstract}

\section{\label{Section1}Introduction}

Remaining within the framework of the standard four-di\-men\-sion\-al General Relativity and the Standard Model of particle physics, it still has not been possible to satisfactorily solve a number of fundamental problems such as
the dark energy and the dark matter problem and neither to unify all fundamental interactions into a single theory.
A possible way to settle these problems consists in  
modifying the Theory of General Relativity and among such attempts, nonlinear $f(R)$ theories \cite{SF,Felice,CDeL,NoijiriOdintsov,Clifton,NOJIRI20171} and theories with extra dimensions \cite{BailinLove,OvWesson,Polchinski,Wesson,Maartens} have gained particular popularity. 

It is indeed appealing to study models which combine both approaches, in other words, to consider nonlinear $f(R)$ models in multidimensional space-time. As regards the cosmological aspects, such models were investigated in  \cite{GMZ1,GMZ2,Timur1,Timur2,Timur3,Timur4,Ketov},
where the authors focused mainly on the problems of internal space stabilization, early and late time accelerated expansion of the Universe and dark matter in the form of gravexcitons/radions. To be viable candidates, these models are required to satisfy the gravitational tests carried out in the Solar system, viz., the deflection of light, time delay of radar echoes and perihelion shift. Strictly speaking, the parametrized post-Newtonian (PPN) parameter $\gamma$ obtained for these models must be in concordance with the constraints imposed by such experimental data. For multidimensional models, the astrophysical setting is the most appropriate to study these effects  \cite{1003.5690,1010.5740,1101.3910}, as it is also for the General Relativity (see, e.g., \cite{Landau}). 
In this setting, the static background metric defined on the product manifold $M=M_4\times M_d$ (where $M_4$ describes external four-dimensional flat space-time and $M_d$ corresponds to the $d$-dimensional internal space) is perturbed by a compact gravitating mass and then, the perturbed metric coefficients are investigated in the weak field limit.

\sloppy The above mentioned approach with respect to the multidimensional nonlinear $f(R)$ models was applied in \cite{1104.1456,1112.1539}, where the extra dimensions were toroidally compactified and hence, the internal space was not curved. One of the main features of this model, brought about by nonlinearity, is that the metric coefficients receive correction terms in the form of the Yukawa potential. The corresponding Yukawa mass reads \cite{1104.1456} ${m_{\mathrm{scal}} = \left[-(D-1)f'(R_0)/(2Df''(R_0))+R_0/D\right]^{1/2}}$, where $D$ denotes the number of spatial dimensions, $R_0$ denotes the background value of the scalar curvature ($R_0=0$ in the case of flat background space-time) and the prime denotes differentiation with respect to $R$. Such massive scalar degree of freedom (dubbed scalaron in \cite{scalaron}) is known to be a characteristic feature of the nonlinear models (see, e.g., \cite{Felice,MYuk1,MYuk2,MYuk3,Novak}).  
In papers \cite{1104.1456,1112.1539}, a point-like matter source is assigned dust-like equation of state (EoS) in our external space and some arbitrary EoS parameter $\Omega$ in the internal space. The nonzero negative $\Omega$ (i.e., tension in the internal space) is then found to be the necessary condition for the PPN parameter $\gamma$ to satisfy the experimental constraints. The particular value $\Omega =-1/2$, which is  the black brane/string condition \cite{1010.5740,1101.3910}, presents an interesting case, since gravitating matter sources in the form of black branes/strings do not destroy the stabilization of the internal space \cite{1010.5740,1101.3910,1201.1756}.

As indicated previously, in multidimensional nonlinear $f(R)$ models considered in \cite{1104.1456,1112.1539}, both the internal and external background spaces are flat. Namely, background matter is absent. To make the background curved, one needs to introduce a background perfect fluid. The linear multidimensional models with spherical compactification of the internal space were considered in \cite{1107.3388,1202.2677,1209.4501,1402.1340,ACZ}. It was shown that the background matter responsible for the internal space curvature has the vacuum-like EoS parameter $\bar\omega_0=-1$ in the external  space and some positive  EoS parameter $\bar\omega_1>0$ in the internal space. It is known \cite{GuntherZhuk} that (i) monopole form fields and the Casimir effect can produce such EoS and (ii) this form of matter stabilizes the internal space (see also \cite{1107.3388,1402.1340}).  In linear models with spherical compactification, corrections to the metric coefficients caused by the gravitating mass acquire the form of the Yukawa potential with the Yukawa mass defined by the radius $a$ of the sphere: $m_{\mathrm{rad}}\sim 1/a$. This scalar degree of freedom (so-called gravexcitons/radions \cite{GuntherZhuk,radion}) is a result of the variations in the internal space volume. It was demonstrated in \cite{1107.3388,1402.1340} that this Yukawa mass and the gravexciton/radion mass are, indeed, exactly equal to each other. For the PPN parameter $\gamma$ to be in agreement with experiments, either the Yukawa mass should be sufficiently large, or for some arbitrary Yukawa mass, the EoS parameter should take on the value $\Omega=-1/2$. 

What happens now if we consider nonlinear $f(R)$ models with spherical compactification of the internal space? How do scalar degrees of freedom associated with the nonlinearity of the model and with internal space fluctuations compete with each other? This is the main subject of the present paper. We also suppose that the background perfect fluid is nonlinear, i.e., the background parameters $\bar\omega_0$ and $\bar\omega_1$ of the EoS in the external and internal spaces are not equal to the squared speed of sound in these spaces. The linear model $f(R)=R$ with spherical compactification and nonlinear background perfect fluid was considered in \cite{1906.08214}. Now, we generalize the study for some arbitrary function $f(R)$. First, we obtain the system of linearized equations for perturbations of the metric coefficients. This system has a rather complicated form, but to our surprise, it appears to be exactly solvable for the arbitrary $f(R)$. In the most general case with the condition $f''(R_0)\neq 0$, the correction terms to the metric coefficients are found as a combination of two Yukawa potential terms. The associated Yukawa characteristic masses $\mu_{1,2}$, however, are neither equal to $m_{\mathrm{scal}}$ and $m_{\mathrm{rad}}$,  nor may be expressed as a simple combination of them. Therefore, the relationship between these masses and $\mu_{1,2}$ is investigated in the relevant limiting cases. For completeness, we also  generalize the linear model $f(R)=R+2\kappa\Lambda_6$ previously considered in \cite{1107.3388,1202.2677,1209.4501} to the case of an arbitrary nonlinear background perfect fluid.

The paper is structured as follows. In section~\ref{Sec2}, we describe the background model which is perturbed by the massive gravitating source. Here, we derive the system of linearized Einstein equations for the perturbed metric coefficients. This system of equations is solved for nonlinear models with $f''(R_0)\neq 0$ in section~\ref{Sec3}. Then, another particular case, $f''(R_0)=0$ with nonlinear perfect fluid is investigated in section~\ref{Sec4}. In section~\ref{Sec5}, we study the case $\omega_0=0$ of the  
nonlinear perfect fluid, viz., zero speed of sound in the external space. In the concluding section~\ref{Sec6} we summarize the obtained results. In Appendix we collect the formulas for the perturbations of the  Ricci tensor which we use
to construct the linearized Einstein equations.

\section{\label{Sec2}Basic equations}

It is well known (see, e.g., \cite{GMZ1,GMZ2}) that in the case of $f(R)$ gravity, the Einstein equations take on the form
\ba{1.1} f'(R)R_{ik}-\frac{1}{2}f(R)g_{ik}-[f'(R)]_{;i;k}+g_{ik}[f'(R)]_{;m;n}g^{mn} \,\nn\\
=\kappa T_{ik}\, ,\ea
which is valid for an arbitrary number of space-time dimensions\footnote{It is well known that $f(R)$ theories are equivalent to scalar-tensor gravity (see, e.g., \cite{SF,Felice,CDeL}). Thus, the dynamical equations (e.g., Eq. \rf{1.1}) can also be reformulated accordingly.  In the present paper, however, we do not make use of this equivalence and to compare our findings with the previous results obtained in papers \cite{1104.1456,1112.1539}, proceed within the framework of the original $f(R)$ theory instead.}. In six-dimensional space-time, $\kappa \equiv2S_5\tilde{G}_{6}/c^4 $ with the total solid angle $S_5=2\pi^{5/2}/\Gamma (5/2)=8\pi^2/3$ and the six-dimensional gravitational constant $\tilde G_{6}$. The trace of this equation reads
\be{1.2} f'(R)R-3f(R)+5[f'(R)]_{;m;n}g^{mn}=\kappa T \, , \ee
where $T=g^{mn}T_{mn}$ stands for the trace of the energy-mo\-men\-tum tensor (EMT). From this point on, we fix the number of space-time dimensions to six.

In the background, we consider a factorized six-di\-men\-sion\-al metric
\be{1.3} ds^2=c^2dt^2-dx^2-dy^2-dz^2-a^2(d\xi^2+\sin^2\xi d\eta^2)\, ,\ee
defined on the product manifold $M=M_4\times M_2$ where $M_4$ is the external/our four-dimensional flat space-time and $M_2$ describes the internal two-dimensional space, which is a sphere of radius $a=\const$. For the metric \rf{1.3}, the only nonzero components of the Ricci tensor are $R^{(0)}_{44}=1$, $R^{(0)}_{55}=\sin^2\xi$ and the scalar curvature $R_0=-2/a^2$. It is worth mentioning that the minus sign in the latter formula follows from the metric adopted here as well as the sign convention for curvature (as in the book \cite{Landau}).

We suppose that the metric \rf{1.3} corresponds to the background matter with the EMT
\be{1.4} \left(T^{i}_{k}\right)^{(0)}=\mathrm{diag}(\bar{\varepsilon},-\bar{p}_0,-\bar{p}_0,-\bar{p}_0,-\bar{p}_1,-\bar{p}_1)\, , \ee

\noindent
where $\bar\varepsilon$ is the background energy density and $\bar p_0 (\bar p_1)$ is the background pressure in the external (internal) space. It is thus this matter that curves the background geometry. Again, at the background level, Eqs.~\rf{1.1} and \rf{1.2} take on the form
\be{1.5} f'(R_0)R^{(0)}_{ik}-\frac{1}{2}f(R_0)g^{(0)}_{ik}=\kappa T^{(0)}_{ik}\, , \ee

\noindent
and
\ba{1.6} f'(R_0)R_0-3f(R_0)= \kappa T^{(0)}\, ,\ea

\noindent
respectively. From Eq.~\rf{1.5}, we obtain the set
\ba{1.7}
-\frac{1}{2}f(R_0)&=&\kappa \bar{\varepsilon}\, , \\
\label{1.8} \frac{1}{2}f(R_0)&=&\kappa\bar{p}_0\, ,\\
\label{1.9} f'(R_0)+\frac{a^2}{2}f(R_0)&=&\kappa\bar{p}_1a^2\, ,
\ea

\noindent
which implies the following definition for the background EMT:

\be{1.10} T^{(0)}_{ik} = \left\{
\begin{array}{ll}
	g^{(0)}_{ik}\bar\varepsilon & \for i,k=0,..,3\, ; \\
	-g^{(0)}_{ik}\bar p_1 & \for i,k=4,5 \, .
\end{array}
\right. \ee
These equations demonstrate that the background EoS parameter in the external space $\bar\omega_0\equiv \bar p_0/\bar\varepsilon =-1$. In order to get the expression for the background EoS parameter $\bar\omega_1\equiv \bar p_1/\bar\varepsilon$ in the internal space, the form of $f(R)$ should be specified. Plugging in \rf{1.7}-\rf{1.10}, one may also verify that Eq.~\rf{1.6} is immediately satisfied. 

Now, we perturb the background model by a static point-like massive source smeared over the internal space
with mass density $\hat\rho(r)=m\delta(\mathbf{r})/V_{\mathrm{int}}$,
where $r=|\mathbf{r}|=\sqrt{x^2+y^2+z^2}$ and the internal space volume $V_{\mathrm{int}}=4\pi a^2$. Since we suppose that this gravitating mass simulates an ordinary astrophysical object, e.g., our Sun, with pressure much less than the energy density, it has the dust-like EoS in the external space. Pressure in the internal space, however, is arbitrary with the EoS parameter $\Omega$. The only nonzero components of the EMT are, then,
$\hat{T}^0_0\approx\hat{\rho} c^2$ and
$\hat{T}^i_k\approx-\delta^i_k\Omega\hat{\rho} c^2$ for $i,k=4,5$. Such type of perturbation preserves the block-diagonal form of the perturbed metric \cite{1107.3388,ACZ}:
\ba{1.11} ds^2&=&\left[1+A^1(x,y,z)\right]c^2dt^2\,\nn\\
&&-\left[1-B^1(x,y,z)\right](dx^2+dy^2+dz^2)\,\nn\\
&&-\left[a^2-G^1(x,y,z)\right](d\xi^2+\sin^2\xi d\eta^2) \, , \ea
where $A^1, B^1$ and $G^1$ denote the first-order corrections to the metric coefficients. In what follows, we consider the perturbed values
$R_{ik}$, $g_{ik}$ and $T_{ik}$ in the form $A_{ik}=A_{ik}^{(0)}+\nobreak A_{ik}^{(1)}$ with consecutive terms denoting the
background values and first-order perturbations, respectively. By the same reasoning, the perturbed scalar curvature  is decomposed as $R=\nobreak R_0+R_1$. As for the function $f(R)$, we get 
\begin{eqnarray}
f(R)&=&f(R_0)+f'(R_0)R_1+O(R_1^2)\, , \label{1.12}\\
f'(R)&=&f'(R_0)+f''(R_0)R_1+O(R_1^2)\, , \label{1.13}
\end{eqnarray}
and Eqs.~\rf{1.1} and \rf{1.2} accordingly yield
\ba{1.14} && f'(R_0)\left(R^{(1)}_{ik}-\frac{1}{2}g^{(0)}_{ik}R_1\right)\,\nn\\
&&-\frac{1}{2}f(R_0)g^{(1)}_{ik}+f''(R_0)R^{(0)}_{ik}R_1\,\nn\\
&&+ \left[g^{(0)}_{ik}(R_1)_{;m;n}g^{(0)mn}-(R_1)_{;i;k}\right]f''(R_0)=\kappa T^{(1)}_{ik} 
 \, , \ea
\ba{1.15} -2f'(R_0)R_1+f''(R_0)\left[R_0R_1+5(R_1)_{;m;n}g^{(0)mn}\right]\,\nn\\
=\kappa T^{(1)}\, .\ea
Here, $ T^{(1)}_{ik}$ and $T^{(1)}$ denote the first-order perturbations of the total EMT and its trace, respectively. Again, up to the first order, the EMT of the perturbed background matter reads
\ba{1.16} \tilde{T}^{i}_{k} \approx \left\{
\begin{array}{lll}
	(\bar{\varepsilon}+\delta\varepsilon)\delta^i_k & \mathrm{for}& i,k=0; \\
	-(\bar{p}_0+\delta p_0)\delta^i_k & \mathrm{for}& i,k=1,2,3; \\
	-(\bar{p}_1+\delta p_1)\delta^i_k & \mathrm{for}& i,k=4,5 \, 
\end{array}
\right. \equiv \left(T^{i}_{k}\right)^{(0)} +\delta T^i_k , \,\nn\\
\ea
indicating
\be{1.17}   
\left(T^i_k\right)^{(1)}= \delta T^i_k+\hat T^i_k\, .
\ee
Let us also assume that
\be{1.18} \delta p_0=\omega_0\delta\varepsilon \, ,\quad \delta p_1=\omega_1\delta \varepsilon\, ,\ee
where $\omega_0$ and $\omega_1$ are constant parameters, which are, generally speaking, not connected with the background quantities: $\omega_0\neq \bar\omega_0$ and $\omega_1\neq\bar\omega_1$. Background matter is thus considered to be some nonlinear perfect fluid and the ratio $\delta p/\delta\varepsilon$, referred to as the squared speed of sound \cite{SoS}, is not equal to the background EoS parameters in the internal and external spaces. 

In this setting, we now turn to the Einstein equations \rf{1.14} and \rf{1.15}, taking into account the expressions for the perturbed Ricci tensor components presented in Appendix. Starting with  Eq.~\rf{1.14}, from the $00$-component we obtain
\be{1.19}
\frac{f'(R_0)}{2}\left(\Delta_3A^1-R_1\right)-\left(\Delta_3 R_1\right)f''(R_0)=\kappa\left(
\delta\varepsilon +\hat\rho c^2\right)\, ,
\ee

\noindent
where we have also used Eq.~\rf{1.7}. The $11$-, $22$- and $33$-com\-po\-nents yield

\ba{1.20}
&&\frac{f'(R_0)}{2}\left[\Delta_3B^1+\frac{\partial^2}{(\partial x^i)^2}\left(-A^1+B^1+2\frac{G^1}{a^2} \right)+R_1\right]\,\nn\\
&&-\frac{1}{2}f(R_0)B^1+\left[\Delta_3 R_1-\frac{\partial^2 R_1}{(\partial x^i)^2}\right]f''(R_0)\nn\\
&&=\kappa\left(-\bar{p}_0B^1+\delta p_0\right)\, , \quad i=1,2,3\, ,
\ea
and using Eq.~\rf{1.9}, the components $44$ and $55$ take on the form
\ba{1.21}
&&\frac{f'(R_0)}{2}\left(\Delta_3 \frac{2G^1}{a^2}+2R_1\right)+f''(R_0)\frac{2R_1}{a^2}+2 \left(\Delta_3 R_1\right)f''(R_0)\, \nn\\
&&=-f'(R_0)\frac{2G^1}{a^4}+2\kappa\left(\delta p_1+\Omega\hat{\rho}c^2\right). 
\ea
The mixed $12$-, $13$- and $23$-components read
\be{1.22}
\frac{1}{2} f'(R_0)\left(-A^1+B^1+2\frac{G^1}{a^2}\right)_{xy}-f''(R_0)\frac{\partial^2 R_1}{\partial x \partial y} =0 \, , 
\ee
\be{1.23} \frac{1}{2} f'(R_0)\left(-A^1+B^1+2\frac{G^1}{a^2}\right)_{xz}-f''(R_0)\frac{\partial^2 R_1}{\partial x \partial z} =0  \, , \ee
\be{1.24} \frac{1}{2} f'(R_0)\left(-A^1+B^1+2\frac{G^1}{a^2}\right)_{yz}-f''(R_0)\frac{\partial^2 R_1}{\partial y \partial z} =0  \, , \ee
and in fact, translate into the relation
\be{1.25} \frac{1}{2} f'(R_0)\left(-A^1+B^1+2\frac{G^1}{a^2}\right)-f''(R_0)R_1=0 \, ,\ee
once we employ the boundary condition that the metric coefficients $A^1, B^1$ and $G^1$, as well as the scalar curvature $R_1$, should tend to zero in the limit $r\to \infty$. We note here that in the linear model with $f''(R_0)= 0$, \rf{1.25} is reduced to its counterpart obtained in papers \cite{1107.3388,1202.2677}.

Using this relation and Eq.~\rf{1.8}, we re-express \rf{1.20} as
\ba{1.26}
\frac{f'(R_0)}{2}\left(\Delta_3B^1+R_1\right)+
\left(\Delta_3 R_1\right)f''(R_0) =\kappa\delta p_0 \, .
\ea

Combining Eqs.~\rf{1.19}, \rf{1.21}, \rf{1.26} and again, employing the relation \rf{1.25}, we obtain
\ba{1.27} &&5f''(R_0)(\Delta_3R_1)+2f'(R_0)R_1+\frac{2}{a^2}f''(R_0)R_1\nn\\
&=&-f'(R_0)\frac{2G^1}{a^4}+\kappa\left[-\delta\varepsilon+\delta p_0+2\delta p_1-\hat{\rho}c^2(1-2\Omega)\right]\, .\nn\\
 \ea
On the other hand, the trace equation \rf{1.15} reads 
 \ba{1.28}
&-&2f'(R_0)R_1+f''(R_0)\left(R_0R_1-5\Delta_3R_1\right)\,\nn\\
&&=\kappa\left[\delta\varepsilon-3\delta p_0-2\delta p_1+\hat{\rho }c^2(1-2\Omega)\right] \, ,
\ea
and added to \rf{1.27}, reveals the relation between $G^1$ and $\delta p_0$:
\be{1.29} 
\delta p_0=-\frac{f'(R_0)}{\kappa a^4}G^1\, . \ee

Before going further, we make two remarks. First, in the case $f''(R_0)=0$, $f'(R_0)=1$ (corresponding also to the linear model $f(R)=R+2\kappa\Lambda_6$) and for linear background perfect fluid characterized by $\omega_0=\bar\omega_0$, $\omega_1=\bar\omega_1$, Eqs.~\rf{1.19}, \rf{1.21}, \rf{1.26} and \rf{1.28} are reduced to the system of equations (22)-(24) investigated earlier in \cite{1202.2677}. Therefore, in what follows, we demand $f''(R_0)\neq 0$ except for the two particular cases considered in sections \ref{Sec4} and \ref{Sec5.2}. Second, given the form of the above equations, we should also demand that $f'(R_0)\neq 0$. In fact, even the tighter constraint of $f'(R_0)> 0$ must be imposed so that the graviton is not a ghost \cite{Felice,Starobinsky}. 

Now, we have a system of Eqs.~\rf{1.19}, \rf{1.21} and \rf{1.26} as well as relations \rf{1.25} and \rf{1.29}. From Eq.~\rf{1.29} and definitions \rf{1.18}, we get
\ba{1.30} \delta \varepsilon=-\frac{f'(R_0)}{\kappa a^4}\frac{1}{\omega_0}G^1\, ,\enspace \delta p_1=-\frac{f'(R_0)}{\kappa
	a^4}\frac{\omega_1}{\omega_0}G^1\, ,\enspace 
\omega_0\neq0\, ,\,\nn\\
\ea
which, when substituted into \rf{1.19}, \rf{1.21} and \rf{1.26} together with the expression for $\delta p_0$ from \rf{1.29}, yield
\ba{1.31}
\frac{f'(R_0)}{2}\Delta_3B^1&+&\frac{f'(R_0)}{2}R_1+f''(R_0)\Delta_3 R_1+\frac{f'(R_0)}{a^4}G^1=0\, ,\nn\\ \ea
\ba{1.32} 
\frac{f'(R_0)}{2}\Delta_3A^1-\frac{f'(R_0)}{2}R_1-f''(R_0)\Delta_3 R_1&+&\frac{f'(R_0)}{a^4}\frac{1}{\omega_0}G^1\,\nn\\
&=&\kappa\hat\rho c^2 \, ,\ea
\ba{1.33}
\frac{f'(R_0)}{2a^2}\Delta_3 G^1&+&\frac{f'(R_0)}{2}R_1+\frac{f''(R_0)}{a^2}R_1+ f''(R_0)\Delta_3 R_1\,\nn\\
&+&\frac{f'(R_0)}{a^4}\left(1+\frac{\omega_1}{\omega_0}\right)G^1= \kappa\Omega\hat{\rho}c^2 \, .\ea
In the next steps, we proceed to solve the system \rf{1.31}-\rf{1.33} together with the relation \rf{1.25} for the functions $A^1$, $B^1$, $G^1$ and $R_1$.

\section{\label{Sec3}Generic model: $f^{\prime\prime}(R_0)\neq 0, \,  \omega_0\neq 0$}

Throughout this section, we study the case of $f^{\prime\prime}(R_0)\neq 0$ (where the linear model with respect to the function $f(R)$ is excluded) for arbitrary values of
the EoS parameters $\omega_0$ and $\omega_1$. In general, these parameters 
are not equal to the background values $\bar\omega_0$ and $\bar\omega_1$, hence, we assume a nonlinear perfect fluid. Additionally, in agreement with Eq.~\rf{1.30}, the equations to be solved are valid only for $\omega_0\neq 0$.

In attempt to solve the set \rf{1.31}-\rf{1.33} and \rf{1.25} for the functions $A^1$, $B^1$, $G^1$ and $R_1$, we make use of the Fourier transform  
\be{2.1}
f(x,y,z)=(2\pi)^{-3/2}\int_{\mathbb{R}^3} d\mathbf{k} e^{i\mathbf{kr}}\tilde{f}(\mathbf{k}) \, .
\ee

Provided also that the previously introduced delta-shaped matter source is smeared over the internal space, i.e., $\hat{\rho}=m\delta(\mathbf{r})/V_{\mathrm{int}}$, the resulting equations read
\ba{2.2}
&-&\frac{f'(R_0)}{2}k^2\tilde{B}^1+\frac{f'(R_0)}{2}\tilde{R}_1-f''(R_0)k^2\tilde{R}_1+\frac{f'(R_0)}{a^4}\tilde{G}^1  \, \nn\\
&&=0\, ,\nn\\
&-&\frac{f'(R_0)}{2}k^2\tilde{A}^1-\frac{f'(R_0)}{2}\tilde{R}_1+f''(R_0)k^2 \tilde{R}_1+\frac{f'(R_0)}{a^4}\frac{1}{\omega_0}\tilde{G}^1\, \nn\\
&&=\kappa\frac{m\tilde{\delta}(\mathbf{k})}{V_{\mathrm{int}}} c^2 \, ,\nn\\
&-&\frac{f'(R_0)}{2a^2}k^2 \tilde{G}^1+\frac{f'(R_0)}{2}\tilde{R}_1+\frac{f''(R_0)}{a^2}\tilde{R}_1- f''(R_0)k^2 \tilde{R}_1\, \nn\\
&&+\frac{f'(R_0)}{a^4}\left(1+\frac{\omega_1}{\omega_0}\right)\tilde{G}^1= \kappa\Omega\frac{m\tilde{\delta}(\mathbf{k})}{V_{\mathrm{int}}}c^2 \, , \nn\\ 
&&\frac{1}{2} f'(R_0)\left(-\tilde{A}^1+\tilde{B}^1+2\frac{\tilde{G}^1}{a^2}\right)-f''(R_0)\tilde{R}_1=0 \, .
\ea

Given the relation $\tilde{\delta}(\mathbf{k})=(2\pi)^{-3/2}$, the solutions $\tilde{A}^1(k)$ and $\tilde{B}^1(k)$ in the Fourier space are

\ba{2.3}
\tilde{A}^1(k)=\frac{(\alpha_{1A}/\alpha_{4})k^4+(\alpha_{2A}/\alpha_{4})k^2+(\alpha_{3}/\alpha_{4})}{k^2\left[k^4+(\alpha_{5}/\alpha_{4})k^2+(\alpha_{6}/\alpha_{4})\right]}\, , \nn\\
\tilde{B}^1(k)=\frac{(\alpha_{1B}/\alpha_{4})k^4+(\alpha_{2B}/\alpha_{4})k^2+(\alpha_{3}/\alpha_{4})}{k^2\left[k^4+(\alpha_{5}/\alpha_{4})k^2+(\alpha_{6}/\alpha_{4})\right]} 
\, ,
\ea

\noindent
where we have introduced the set of coefficients
\ba{2.4} 
\alpha_{1A}&\equiv&\kappa^{\prime}\left[4f''(R_0)a^4\omega_0(2+\Omega) \right]\, ,\nn\\
\alpha_{1B}&\equiv&\kappa^{\prime}\left[2f''(R_0)a^4\omega_0(1-2\Omega)\right]\, , \nn\\
\alpha_{2A}&\equiv&\kappa^{\prime}\left\lbrace 4f''(R_0)a^2\left[\Omega (2+\omega_0)-2\omega_1-2\omega_0\right]\, \right.\nn\\
&-&\left.f'(R_0)a^4\omega_0(3+2\Omega)\right\rbrace \, ,\nn\\
\alpha_{2B}&\equiv&\kappa^{\prime}\left\lbrace 4f''(R_0)a^2\left[\Omega (1+2\omega_0)-\omega_1\right]\,\right.\nn\\
&-&\left.f'(R_0)a^4\omega_0(1-2\Omega)\right\rbrace \, ,\nn\\
\alpha_{3}&\equiv&-\kappa^{\prime}\left\{2f'(R_0)a^2\left[\Omega(1+\omega_0)-\omega_1\right]+4f''(R_0)\omega_0 \right\} \, ,\nn\\
\ea
and
\ba{2.5}
\alpha_{4}&\equiv&-5f''(R_0)f'(R_0)a^4\omega_0\, , \nn\\
\alpha_{5}&\equiv&2a^2f'(R_0)\left[a^2f'(R_0)\omega_0+f''(R_0)\left(1+3\omega_0+3\omega_1\right)\right] \, , \nn\\
\alpha_{6}&\equiv&f'(R_0)\,\nn\\
&&\times\left[2f''(R_0)(\omega_0-1)-a^2f'(R_0)(1+\omega_0+2\omega_1)\right] \, ,\nn\\
\ea
together with $\kappa'\equiv(2\pi)^{-3/2} V^{-1}_{\mathrm{int}} \kappa mc^2$. For $\tilde{G}^1(k)$ and $\tilde{R}_1(k)$, the solutions have the form  
\ba{2.6}
\tilde{G}^1(k) =\frac{(\alpha_{1G}/\alpha_4)k^2+(\alpha_{2G}/\alpha_4)}{k^4+(\alpha_{5}/\alpha_{4})k^2+(\alpha_{6}/\alpha_{4})} \, ,\nn \\ 
\tilde{R}_1(k)=
\frac{(\alpha_{1R}/\alpha_4)k^2+(\alpha_{2R}/\alpha_4)}{k^4+(\alpha_{5}/\alpha_{4})k^2+(\alpha_{6}/\alpha_{4})}  \, .\ea
The coefficients $\alpha_{1G}$ and $\alpha_{1R}$ may be expressed in terms of $\alpha_{1A}$ and $\alpha_{1B}$ as
\be{2.7}
\alpha_{1G} \equiv \left(\frac{\alpha_{1A}}{2}-\alpha_{1B}\right)a^2, \quad \alpha_{1R} \equiv -\frac{f'(R_0)}{f''(R_0)}\left(\frac{\alpha_{1B}}{2} \right) \, ,
\ee
and the remaining two coefficients in \rf{2.6} read
\ba{2.8}
\alpha_{2G}&\equiv&-\kappa^{\prime}\left[ 2f''(R_0)a^4\omega_0+f'(R_0)a^6\omega_0(1+2\Omega)\right] \, ,\nn\\
\alpha_{2R}&\equiv&\kappa^{\prime}\left\{2f'(R_0)a^2\left[ \Omega(\omega_0-1)+\omega_0+\omega_1\right]\right\} \, .  
\ea

It is worth noting that the conditions $f'(R_0)\neq 0$,  $f''(R_0)\neq 0$, and $\omega_0\neq 0$ herein require $\alpha_4\neq 0$.
\subsection{Different roots:  $\mu_{1},\mu_{2}>0$, $\mu_1\neq\mu_2$}

In this subsection we consider the case when the polynomial
\be{2.9}
k^4+(\alpha_{5}/\alpha_{4})k^2+(\alpha_{6}/\alpha_{4}) = \left(k^2+\mu^2_1\right)\left(k^2+\mu^2_2\right) \, ,
\ee
has two distinct negative roots $k^2=-\mu_1^2$ and $k^2=-\mu_2^2$. Given the quadratic form of these roots, without loss of generality, we may choose to proceed with the positive values: $\mu_{1},\mu_{2}>0$ . 

After some algebra, the functions \rf{2.3} and \rf{2.6} can be cast into the form
\ba{2.10}
\tilde{A}^1(k)&=&\frac{\beta_{1A}}{k^2+\mu_1^2}+\frac{\beta_{2A}}{k^2+\mu_2^2}+\frac{\beta_{3}}{k^2} \, ,\nn\\
\tilde{B}^1(k)&=&\frac{\beta_{1B}}{k^2+\mu_1^2}+\frac{\beta_{2B}}{k^2+\mu_2^2}+\frac{\beta_{3}}{k^2} \, ,
\ea
and
\ba{2.11}
\tilde{G}^1(k)&=&\frac{\beta_{1G}}{k^2+\mu_1^2}+\frac{\beta_{2G}}{k^2+\mu_2^2} \, ,\nn \\ 
\tilde{R}_1(k)&=&\frac{\beta_{1R}}{k^2+\mu_1^2}+\frac{\beta_{2R}}{k^2+\mu_2^2} \, , 
\ea
with the coefficients
\ba{2.12} 
\beta_{1A(B)}&\equiv&\frac{(\alpha_{1A(B)}/\alpha_4)\mu_1^2-\alpha_{2A(B)}/\alpha_4+(\alpha_{3}/\alpha_4)\mu_1^{-2}}{\mu_1^2-\mu_2^2}\, ,\nn\\
\beta_{2A(B)}&\equiv&-\frac{(\alpha_{1A(B)}/\alpha_4)\mu_2^2-\alpha_{2A(B)}/\alpha_4+(\alpha_{3}/\alpha_4)\mu_2^{-2}}{\mu_1^2-\mu_2^2}\, , \nn\\
\beta_{3}&\equiv&(\alpha_{3}/\alpha_4)\mu_1^{-2}\mu_2^{-2}\, ,\nn\\
\beta_{1G(R)}&\equiv&\frac{(\alpha_{1G(R)}/\alpha_4)\mu_1^2-\alpha_{2G(R)}/\alpha_4}{\mu_1^2-\mu_2^2} \, ,\nn\\
\beta_{2G(R)}&\equiv&-\frac{(\alpha_{1G(R)}/\alpha_4)\mu_2^2-\alpha_{2G(R)}/\alpha_4}{\mu_1^2-\mu_2^2} \, .
\ea

The terms of the form of $1/k^2$ and $1/(k^2+\mu^2)$ in \rf{2.10} and \rf{2.11} correspond to the Newtonian and Yukawa potentials, respectively. Therefore, in the position space we obtain
\ba{2.13}
A^1(r)=\sqrt{\frac{\pi}{2}}\frac{1}{r}&&\left[\beta_3+\beta_{1A}\exp\left(-\mu_1r\right)+\beta_{2A}\exp\left(-\mu_2r\right)\right]\,,\nn\\
\\
B^1(r)=\sqrt{\frac{\pi}{2}}\frac{1}{r}&&\left[\beta_3+\beta_{1B}\exp\left(-\mu_1r\right)+\beta_{2B}\exp\left(-\mu_2r\right)\right] \, ,\nn\\
\label{2.14}\\
\label{2.15}
G^1(r)=\sqrt{\frac{\pi}{2}}\frac{1}{r}&&\left[\beta_{1G}\exp\left(-\mu_1r\right)+\beta_{2G}\exp\left(-\mu_2r\right)\right] \, ,\\
\label{2.16}
R_1(r)=\sqrt{\frac{\pi}{2}}\frac{1}{r}&&\left[\beta_{1R}\exp\left(-\mu_1r\right)+\beta_{2R}\exp\left(-\mu_2r\right)\right] \, ,
\ea

\noindent
consistent with the zero boundary condition for $|\mathbf{r}|\to\infty $. Parameters $\mu_{1,2}^{-1}$ define the characteristic ranges of the Yukawa interaction. For the polynomial \rf{2.9}, we get

\be{2.17}
\mu_{1,2}^2=\frac{1}{2}\left(\frac{\alpha_5}{\alpha_4}\pm\frac{\sqrt{\alpha_5^2-4\alpha_4\alpha_6}}{\alpha_4}\right) \, ,
\ee

\noindent
which, upon substitution of the explicit expressions in \rf{2.5}, reads
\ba{2.18}
&{}&\mu_{1,2}=\frac{1}{\sqrt{5}}\frac{1}{a}\bm\left[-a^2\frac{f'(R_0)}{f''(R_0)}-\frac{(1+3\omega_0+3\omega_1)}{\omega_0} \right.\, \nn\\
&&\mp\frac{1}{f''(R_0)\omega_0}\,\nn\\
&&\times\left[a^4f'^2(R_0)\omega_0^2+a^2f'(R_0)f''(R_0)(-3+\omega_0-4\omega_1)\omega_0\right.\,\nn\\\
&&+\left.\left.f''^2(R_0)\right.\right.\,\nn\\
&&\left.\left.\times\left(19\omega_0^2+2\omega_0(-2+9\omega_1)+\left(1+3\omega_1\right)^2\right)\right]^{1/2}\bm\right]^{1/2} .
\ea

In the limiting case $r\to\infty$, $A^1$ and $B^1$ take on the same value
\ba{2.19}
&&A^1(r\rightarrow\infty)=B^1(r\rightarrow\infty) = \sqrt{\frac{\pi}{2}}\frac{1}{r}\beta_3\,\nn\\
&&=
\sqrt{\frac{\pi}{2}}\frac{1}{r}\frac{\alpha_3}{\alpha_6}\equiv
 -\sqrt{\frac{\pi}{2}}\frac{1}{r}\kappa^{\prime}\nu = -\frac{2}{c^2}\frac{1}{4\pi}\frac{S_D\tilde{G}_{\mathcal{D}}}{V_{\mathrm{int}}}\frac{m}{r}\nu \, ,
\ea

\noindent
where

\ba{2.20}
\nu\equiv\frac{2f'(R_0)a^2\left[\Omega(1+\omega_0)-\omega_1\right]+4f''(R_0)\omega_0}{f'(R_0)\left[2f''(R_0)(\omega_0-1)-a^2f'(R_0)(1+\omega_0+2\omega_1)\right]} .\,\nn\\ 
\ea

It is well known (see, e.g., \cite{Landau}) that the metric coefficient $A^1$ defines the gravitational potential and from Eq.~\rf{2.19}, we see that it tends to the Newtonian potential in the corresponding limit $r\to\infty$. With this regard, demanding $A^1(r\to\infty)=2\varphi_N/c^2$ for $\varphi_N=-G_N m/r$, we deduce the relation between the multidimensional gravitational constant $\tilde{G}_{\mathcal{D}}$ and the Newtonian gravitational constant $G_N$ in the form

\be{2.21}\frac{S_D\tilde{G}_{\mathcal{D}}}{V_{\mathrm{int}}}\nu=
4\pi G_N\, .
\ee

\noindent
For $\omega_0=\bar\omega_0=-1$, the expression \rf{2.20} for $\nu$ is significantly simplified to yield

\be{2.22}
\frac{S_D\tilde{G}_{\mathcal{D}}}{V_{\mathrm{int}}}\frac{1}{f^{\prime}(R_0)}=4\pi G_N \, ,
\ee

\noindent
which clearly demands the positiveness of $f^{\prime}(R_0)$. 

\subsubsection*{Example: $f(R)=R+\xi R^2$}

In order to obtain specific estimates, it is necessary to determine the form of the function $f(R)$ and to this end, we consider the popular 
quadratic model ${f(R)=R+\xi R^2}$. Herein, imposing the condition $f''(R_0)\neq 0$ immediately results in $\xi\neq0$. Requiring $f'(R_0)>0$ as well, we obtain the inequality ${a^2>4\xi}$, which is satisfied for all negative values of $\xi$. 

For such choice, the Yukawa parameters $\mu_{1,2}$ \rf{2.18} read
\ba{2.23}
&&\mu_{1,2}=\frac{1}{\sqrt{10}}\frac{1}{a}\left[-\frac{a^2}{\xi}-\frac{2(1+\omega_0+3\omega_1)}{\omega_0}\,\right.\nn\\
&&\left.\mp\frac{1}{\xi\omega_0}\left[ a^4\omega_0^2-2a^2\xi(3+3\omega_0+4\omega_1)\omega_0\right.\right.\,\nn\\\
&&+\left.\left.4\xi^2\left(21\omega_0^2+\omega_0(2+26\omega_1)+\left(1+3\omega_1\right)^2\right)\right]^{1/2}\right]^{1/2}, \,\nn\\
\ea
and for $\nu$ \rf{2.20}, we get
 \makeatletter 
 \def\@eqnnum{{\normalsize \normalcolor (\theequation)}} 
  \makeatother
  
{\small
\ba{2.24}
\nu=-\frac{2a^2\left\{4\xi[\omega_0+\omega_1-\Omega(1+\omega_0)]+a^2[-\omega_1+\Omega(1+\omega_0)]\right\}}{(a^2-4\xi)\left[a^2(1+\omega_0+2\omega_1)-8\xi(\omega_0+\omega_1) \right]} \, .\,\nn\\
\ea}

\noindent
From Eq.~\rf{2.23}, we see that the Yukawa parameters $\mu_{1,2}$ are not related to the gravexciton/radion mass

\be{2.24b}
m_{\mathrm{rad}}\sim 1/a \, ,
\ee

\noindent
and to the scalaron mass \cite{1104.1456}

\ba{2.24a}
m_{\mathrm{scal}}&=& \frac{1}{\sqrt{5}}\left(-\frac{2f'(R_0)}{f''(R_0)}+R_0\right)^{1/2}\,\nn\\
&=&\frac{1}{\sqrt{5|\xi|}} \left(\frac{2|\xi|}{a^2}-\frac{|\xi|}{\xi}\right)^{1/2} \, 
\ea

\noindent
by simple expressions. Nevertheless, if ${|\xi|\sim a^2}$, all masses are of the order of $1/a$: $m_{\mathrm{scal}}\sim m_{\mathrm{rad}}\sim \mu_{1,2}\sim 1/a$ (up to the natural assumption $\omega_0,\omega_1\sim O$(1)), so it appears interesting to consider a number of limiting cases to further explore such possible connections. 

First, we study the case $|\xi|\gg a^2$. It can easily be seen that in this limit, the scalaron mass $m_{\mathrm{scal}}\sim 1/a$ (for any sign of $\xi$) is similar to the radion mass $m_{\mathrm{rad}}$. Moreover, the parameters $\mu_{1,2}$ read
\ba{2.25}
&&\mu_{1,2}\approx\frac{1}{\sqrt{5}}\frac{1}{a}\left[-\frac{(1+\omega_0+3\omega_1)}{\omega_0}\right.\,\nn\\
&&\left.\mp\frac{|\xi|}{\xi}\frac{1}{\omega_0}\left[21\omega_0^2+(2+26\omega_1)\omega_0+(1+3\omega_1)^2\right]^{1/2}\right]^{1/2} \, .\,\nn\\ 
\ea  

\noindent
As we adopt the natural assumption $\omega_0,\omega_1\sim O(1)$, it is possible to see that both parameters $\mu_{1,2}$ as well as the scalaron and radion masses are defined by the radius $a$ of the sphere:  

\be{2.25a}
\mu_{1,2}\sim m_{\mathrm{scal}}
\sim m_{\mathrm{rad}}
\sim \frac{1}{a}\, .
\ee

\noindent
Obviously, for $r\gg a$, the exponential terms in \rf{2.13} and \rf{2.14} can be dropped and for the parameter $\nu$ \rf{2.24}, we find

\be{2.26}
\nu=\frac{a^2\left[\Omega(1+\omega_0)-\omega_0-\omega_1\right]}{4\xi(\omega_0+\omega_1)} \, .
\ee

\noindent
Choosing $\omega_0=\bar\omega_0=-1$, the above expression implies
\be{2.27}
-\frac{S_D\tilde{G}_{\mathcal{D}}}{V_{\mathrm{int}}}\frac{a^2}{4\xi}=4\pi G_N \, , \quad \xi<0 \, ,
\ee

\noindent
which indeed corresponds to \rf{2.22} since $R_0=-2/a^2$ and again, $|\xi|\gg a^2$.
Keeping in mind that $V_{\mathrm{int}}=4\pi a^2$, we realize that in this limit and for the particular case of $\omega_0=-1$, the relation between gravitational constants becomes independent of the internal space radius $a$ (up to some $O(a^2)$ correction term).

Now, we consider the reverse limiting case $|\xi|\ll a^2$. For the scalaron mass, we have $m_{\mathrm{scal}}\sim \left(-1/\xi\right)^{1/2}$, which is valid for negative values of $\xi$. For the parameters $\mu_{1,2}$ we obtain
\ba{2.27a}
\mu_{1,2} &\approx& 
\frac{1}{\sqrt{10}}\frac{1}{a}\,\nn\\
&&\times\left[-\frac{a^2}{\xi}\mp\frac{|\omega_0|}{\omega_0}\frac{a^2}{\xi}\left(1-\frac{\xi}{a^2}\frac{3+3\omega_0+4\omega_1}{\omega_0}\right)\right]^{1/2}\, ,\nn\\
\ea

\noindent 
which, for $\omega_0<0$, are reduced to the following:

\ba{2.27b}
\mu_1 &\approx& \frac{1}{\sqrt{10}}\frac{1}{a}\left(-\frac{3+3\omega_0+4\omega_1}{\omega_0}\right)^{1/2}
\sim \frac{1}{a}\sim m_{\mathrm{rad}}\, ,\\
\label{2.27c}
\mu_2 &\approx& \frac{1}{\sqrt{5}}\frac{1}{a}\left(-\frac{a^2}{\xi}\right)^{1/2}
= \frac{1}{\sqrt{5|\xi|}}\approx m_{\mathrm{scal}}\, .
\ea

\noindent
As for the important particular value $\omega_0=-1$, the inequality $3+3\omega_0+4\omega_1>0$ following from  \rf{2.27b} leads to the condition $\omega_1>0$, similar to what has been obtained in \cite{1107.3388,1202.2677,1402.1340}. On the other hand, Eq.~\rf{2.27c} requires $\xi<0$. Since $a\gg \sqrt{|\xi|}$, $\mu_1 \ll \mu_2$ and in Eqs.~\rf{2.13}-\rf{2.16} we may drop the exponential functions with $\mu_2$. Consequently, given the solutions with a single Yukawa potential, we can apply the results of the inverse square law experiments \cite{ISL} to get restrictions on the parameters of the model, similar to the ones we obtain at the very end of section \ref{Sec4}. For example, if $\beta_{1A}/\kappa',\omega_1\sim O(1)$, then $a\lesssim 10^{-3}$cm \cite{ISL} and $\mu_1\gtrsim 10^{-2}$eV. This is, of course, a very rough estimate since the value of $\beta_{1A}$ strongly depends on $\xi$ and $a$. 

Let us now turn to the limit  $a\to\infty$. The limiting values of the Yukawa parameters are found to be 

\be{2.27d}
\mu_1 = 0\, , \quad \mu_2=\sqrt{\frac{-1}{5\xi}} = m_{\mathrm{scal}}  \,   ,
\ee

\noindent
and accordingly, the metric coefficients $A^1(r)$ and $B^1(r)$ have the forms

\ba{2.27e}A^1(r) &=& \sqrt{\frac{\pi}{2}}\frac{1}{r}\kappa'\left[-\frac{2\Omega +3}{2}+\frac{\left(2\Omega-1\right)}{10}\exp\left(-\mu_2 r\right)\right]\, ,\nn\\
\\\label{2.27f}
B^1(r) &=& \sqrt{\frac{\pi}{2}}\frac{1}{r}\kappa'\left[\frac{2\Omega-1}{2}+\frac{\left(1-2\Omega\right)}{10}\exp\left(-\mu_2 r\right)\right] \, .\nn\\ 
\ea

\noindent
As for the remaining corrections, we first renormalize $G^1(r)$ in the perturbed metric \rf{1.11} and replace it by the term $a^2G^1(r)$. Then, the newly defined function in the limit $a\to\infty$ reads

\be{2.27g} 
G^1(r) = \sqrt{\frac{\pi}{2}}\frac{1}{r}\kappa'\left[-\frac{1+2\Omega}{2}+\frac{1-2\Omega}{10}\exp\left(-\mu_2 r\right)\right] \, .\ee

\noindent
Finally, the perturbed scalar curvature follows as   

\be{2.27k}
R_1(r) = \sqrt{\frac{\pi}{2}}\frac{1}{r}\kappa'\left[\frac{\left(1-2\Omega\right)}{10\xi}\exp\left(-\mu_2 r\right)\right].
\ee

\noindent
Expressions \rf{2.27e}-\rf{2.27k} exactly coincide with the formulas in paper \cite{1112.1539} (for $D=5$ and up to the evident substitution $\Omega \to 2 \Omega$) devoted to nonlinear models with toroidal compactification of the internal space.

From \rf{2.27e} and \rf{2.27f}, the condition $A^1(r\to\infty)=B^1(r\to\nobreak\infty)$ requires $\Omega=-1/2$, which is the EoS for black strings, and one consequently obtains 

\be{2.27l}\nu=1 \, ,\quad \frac{S_D\tilde{G}_{\mathcal{D}}}{V_{\mathrm{int}}}=4\pi G_N \, . \ee

\subsection{Equal roots:  $\mu_1 =\mu_2\equiv \mu >0$}

In the case $\mu_1=\mu_2\equiv \mu$, Eqs.~\rf{2.10} and \rf{2.11} can be rewritten as

\ba{2.28}
\tilde{A}^1(k)&=&\frac{\gamma_{1A}}{k^2+\mu^2}+\frac{\gamma_{2A}}{(k^2+\mu^2)^2}+\frac{\gamma_{3}}{k^2} \, ,\nn \\ 
\tilde{B}^1(k)&=&\frac{\gamma_{1B}}{k^2+\mu^2}+\frac{\gamma_{2B}}{(k^2+\mu^2)^2}+\frac{\gamma_{3}}{k^2} \, ,
\ea

\noindent
and
\ba{2.29}
\tilde{G}^1(k)&=&\frac{\gamma_{1G}}{k^2+\mu^2}+\frac{\gamma_{2G}}{(k^2+\mu^2)^2} \, ,\nn\\
\tilde{R}_1(k)&=&\frac{\gamma_{1R}}{k^2+\mu^2}+\frac{\gamma_{2R}}{(k^2+\mu^2)^2} \, , 
\ea

\noindent
where we have introduced the coefficients
\ba{2.30} 
\gamma_{1A(B)}&\equiv&
\frac{\alpha_{1A(B)}\mu^4-\alpha_3}{\alpha_4\mu^4}
\, ,\nn\\
\gamma_{2A(B)}&\equiv&
\frac{\alpha_{2A(B)}\mu^2-\alpha_{1A(B)}\mu^4-\alpha_3}{\alpha_4\mu^2}\, , \nn\\
\gamma_{3}&\equiv&
\frac{\alpha_3}{\alpha_4\mu^4}
\, ,\nn\\
\gamma_{1G(R)}&\equiv&
\frac{\alpha_{1G(R)}}{\alpha_4} 
\, ,\nn\\
\gamma_{2G(R)}&\equiv&
\frac{\alpha_{2G(R)}-\alpha_{1G(R)}\mu^2}{\alpha_4} \, .
\ea

\noindent
The explicit expression for the root now reads

\be{2.31}
\mu=\sqrt{\frac{\alpha_5}{2\alpha_4}}=\sqrt{-\frac{1}{5}\left(\frac{f'(R_0)}{f''(R_0)}+\frac{(1+3\omega_0+3\omega_1)}{a^2\omega_0}\right)} \, .
\ee

\noindent
In contrast with Eqs.~\rf{2.10} and \rf{2.11}, 
where the Newtonian and Yukawa potentials are the only contributions to the metric coefficients, the term $1/(k^2+\mu^2)^2$ in \rf{2.28} and \rf{2.29} provides new contribution in the form of  a pure exponential potential. Therefore, in position space we get

\be{2.32}
A^{1}(r)=\sqrt{\frac{\pi}{2}}\frac{1}{r} \left[\gamma_3 +\left(\gamma_{1A}+\frac{r}{2\mu}\gamma_{2A}\right)\exp\left(-\mu r\right)\right] \, ,
\ee

\be{2.33}
B^{1}(r)=\sqrt{\frac{\pi}{2}}\frac{1}{r} \left[\gamma_3 +\left(\gamma_{1B}+\frac{r}{2\mu}\gamma_{2B}\right)\exp\left(-\mu r\right)\right] \, ,
\ee

\be{2.34}
G^{1}(r)=\sqrt{\frac{\pi}{2}}\frac{1}{r} \left(\gamma_{1G}+\frac{r}{2\mu}\gamma_{2G}\right)\exp\left(-\mu r\right) \, ,
\ee

\be{2.35}
R_{1}(r)=\sqrt{\frac{\pi}{2}}\frac{1}{r} \left(\gamma_{1R}+\frac{r}{2\mu}\gamma_{2R}\right)\exp\left(-\mu r\right) \, .
\ee

\noindent
Taking into account \rf{2.31}, we can easily see that $\gamma_3=\alpha_3/(\alpha_4\mu^4)=\alpha_3/\alpha_6=\beta_3$. Thus, for the limiting case $r\to\nobreak\infty$, we drop the exponential terms in the above set and reproduce Eqs.~\rf{2.19}-\rf{2.22} to conclude that the corresponding physical interpretations therein hold valid also for the model with equal roots. For finite $r$, the Yukawa potential receives a ``corrupted'' prefactor presented as a combination of the $1/r$ term and a second term independent of the three-dimensional distance $r$.

\subsubsection*{Example: $f(R)=R+\xi R^2$}

Substituting $f(R)= R+\xi R^2$ in \rf{2.31}, we obtain

\be{2.36}
\mu=\frac{1}{\sqrt{|\xi|}}\left(-\frac{|\xi|}{a^2}\frac{1+\omega_0+3\omega_1}{5\omega_0}-\frac{1}{10}\frac{|\xi|}{\xi}\right)^{1/2}\, .
\ee

\noindent
Following the reasoning in the previous subsection, we again consider two limiting cases ${|\xi|\gg a^2}$ and ${|\xi|\ll a^2}$, respectively. As we have already seen, for $|\xi|\gg a^2$, the scalaron mass becomes inversely proportional to the internal space radius: $ m_{\mathrm{scal}}\sim 1/a$. In this limit, the parameter $\mu$ reads

\be{2.38}
\mu\approx\frac{1}{a}\left(\frac{1+\omega_0+3\omega_1}{-5\omega_0}\right)^{1/2}
\sim m_{\mathrm{scal}}
\sim m_{\mathrm{rad}}
\sim \frac{1}{a}\,  ,
\ee

\noindent
and all characteristic masses appear to be defined by the radius of the sphere.
Clearly, the argument of the square root must be positive. For example, in the important case $\omega_0=\bar\omega_0=-1$, the EoS parameter $\omega_1$ should be positive analogous to the conclusion following from equation \rf{2.27b}.

In the limit ${|\xi|\ll a^2}$ the scalaron mass $ m_{\mathrm{scal}}\sim 1/\sqrt{|\xi|},\, \xi<0$. For the parameter $\mu$, we obtain

\be{2.36a}
\mu\approx 
\sqrt{\frac{1}{10|\xi|}} \sim m_{\mathrm{scal}}\, ,
\ee

\noindent
and deduce that it is proportional to the scalaron mass.

\section{\label{Sec4}The case $f^{\prime\prime}(R_0)=0, \omega_0\neq 0$}

In this section, we consider the case of a zero value of the second derivative $f''(R_0)=0$. Nevertheless, we demand that the first derivative is still non-vanishing:  $f'(R_0)\neq0$. A particular example of such case is the linear model where $f(R)=R+2\kappa\Lambda_6$. In what follows, the EoS parameters $\omega_0$ and $\omega_1$ are not equal to the background values  $\bar\omega_0$ and $\bar\omega_1$ in general and we assume the additional inequality $\omega_0\neq 0$. Then, as it follows from Eqs.~\rf{2.4}, \rf{2.5}, \rf{2.7}, and \rf{2.8}, coefficients $\alpha_{1A}=\alpha_{1B}=\alpha_{4}=\alpha_{1G}=0$. For the metric coefficients \rf{2.3} and \rf{2.6}, we get

\ba{3.1}
\tilde{A}^1(k)&=& \frac{(\alpha_{2A}/\alpha_5)k^2+(\alpha_{3}/\alpha_5)}{k^2\left[k^2+(\alpha_{6}/\alpha_5)\right]}\, , \nn\\
\tilde{B}^1(k)&=& \frac{(\alpha_{2B}/\alpha_5)k^2+(\alpha_{3}/\alpha_5)}{k^2\left[k^2+(\alpha_{6}/\alpha_5)\right]}  \, ,
\ea

\noindent
and
\ba{3.2}
\tilde{G}^1(k)&=& \frac{(\alpha_{2G}/\alpha_{5})}{k^2+(\alpha_{6}/\alpha_5)}\, ,\nn\\
\tilde{R}_1(k)&=& \frac{(\alpha_{1R}/\alpha_5)k^2+(\alpha_{2R}/\alpha_5)}{k^2+(\alpha_{6}/\alpha_{5})} \, .
\ea
It is also possible to re-express the above solutions in the form of \rf{2.10} and \rf{2.11}. For example, 

\be{3.3}
\tilde{A}^1(k)=\frac{(\alpha_{2A}/\alpha_5)-(\alpha_{3}/\alpha_5)\mu^{-2}}{k^2+\mu^2}+\frac{(\alpha_{3}/\alpha_5)\mu^{-2}}{k^2} \, ,
\ee  

\be{3.4}
\tilde{R}_{1}(k)=\frac{\alpha_{1R}}{\alpha_5}+\frac{(\alpha_{2R}/\alpha_5)-\mu^2(\alpha_{1R}/\alpha_5)}{k^2+\mu^2} \, ,
\ee
where $\mu^2 \equiv \alpha_6/\alpha_5$. Choosing ${\mu>0}$, the solutions in the position space read

\be{3.5}
A^1(r)= \sqrt{\frac{\pi}{2}}\frac{1}{r}\left[\frac{\alpha_3}{\alpha_6}+\left(\frac{\alpha_{2A}}{\alpha_5}-\frac{\alpha_{3}}{\alpha_6}\right)\exp\left(-\mu r\right)\right] \, ,
\ee

\be{3.6}
B^1(r)=\sqrt{\frac{\pi}{2}}\frac{1}{r}\left[\frac{\alpha_3}{\alpha_6}+\left(\frac{\alpha_{2B}}{\alpha_5}-\frac{\alpha_{3}}{\alpha_6}\right)\exp\left(-\mu r\right)\right]\, ,
\ee
\be{3.7}
G^1(r)=\sqrt{\frac{\pi}{2}}\frac{1}{r}\left[\left(\frac{\alpha_{2G}}{\alpha_5}\right)\exp\left(-\mu r\right) \right]\, ,
\ee
\ba{3.8}
R_1(r)&=&(2\pi)^{3/2}\delta(\mathbf{r})\frac{\alpha_{1R}}{\alpha_5}\,\nn\\
&&+\sqrt{\frac{\pi}{2}}\frac{1}{r}\left[\left(\frac{\alpha_{2R}}{\alpha_5}-\frac{\alpha_{1R}}{\alpha_5}\cdot\frac{\alpha_6}{\alpha_5}\right)\exp\left(-\mu r\right) \right] \, .\,\nn\\
\ea
The first term on the right-hand side (RHS) of \rf{3.8} follows from the constant $\alpha_{1R}/\alpha_5$ in \rf{3.4}.

To analyze these expressions, we need to define the first derivative $f'(R_0)$. For this purpose, we consider the significant case of $f'(R_0)=1$, which yields

\ba{3.9}
A^1(r)&=&\sqrt{\frac{\pi}{2}}\frac{1}{r}\kappa^{\prime}\left\lbrace\frac{2\left[\Omega(1+\omega_0)-\omega_1\right]}{(1+\omega_0+2\omega_1)}\right.\,\nn\\
&&-\left[\frac{3+2\Omega}{2}+\frac{2\left[\Omega(1+\omega_0)-\omega_1\right]}{(1+\omega_0+2\omega_1)}\right]\,\nn\\
&&\left.\times\exp\left(-\sqrt{-\frac{1}{2}\frac{(1+\omega_0+2\omega_1)}{\omega_0}} \frac{r}{a}\right)\right\rbrace \, ,
\ea

\ba{3.10}
B^1(r)&=&\sqrt{\frac{\pi}{2}}\frac{1}{r}\kappa^{\prime}\left\lbrace\frac{2\left[\Omega(1+\omega_0)-\omega_1\right]}{(1+\omega_0+2\omega_1)}\right.\,\nn\\
&&-\left[\frac{1-2\Omega}{2}+\frac{2\left[\Omega(1+\omega_0)-\omega_1\right]}{(1+\omega_0+2\omega_1)}\right]\,\nn\\
&&\left.\times\exp\left(-\sqrt{-\frac{1}{2}\frac{(1+\omega_0+2\omega_1)}{\omega_0}} \frac{r}{a}\right)\right\rbrace\, ,
\ea
\ba{3.11}
G^1(r)&=&-\sqrt{\frac{\pi}{2}}\frac{1}{r}\kappa^{\prime}a^2\left(\frac{1}{2}+\Omega\right)\,\nn\\
&&\times\exp\left(-\sqrt{-\frac{1}{2}\frac{(1+\omega_0+2\omega_1)}{\omega_0}} \frac{r}{a}\right) \, ,
\ea

\ba{3.12}
R_1(r)&=&-\kappa^{\prime}(2\pi)^{3/2}\delta(\mathbf{r})\left(\frac{1}{2}-\Omega\right)\,\nn\\
&&+\sqrt{\frac{\pi}{2}}\frac{1}{r}\kappa^{\prime}\left[\frac{(2\omega_1+3\omega_0-1)(1+2\Omega)}{4a^2\omega_0}\,\right.\nn\\
&&\left.\times\exp\left(-\sqrt{-\frac{1}{2}\frac{(1+\omega_0+2\omega_1)}{\omega_0}} \frac{r}{a}\right) \right] \, .
\ea

\noindent
It can be verified that these metric coefficients satisfy Eqs.~\rf{1.25} and \rf{1.31}-\rf{1.33}.

For the specific linear model $f(R)=R+2\kappa\Lambda_6$ with linear background perfect fluid  $\omega_0=\bar\omega_0=-1$ and $\omega_1=\bar\omega_1=\Lambda_6/[1/(\kappa a^2)-\Lambda_6]$ (see Eqs.~\rf{1.7}-\rf{1.9}), the metric coefficients \rf{3.9}-\rf{3.12} exactly coincide with the solutions obtained in paper \cite{1107.3388} (for $\Omega=0$) and in papers \cite{1202.2677,1209.4501} (for arbitrary $\Omega$). Therefore, solutions \rf{3.9}-\rf{3.12} indeed generalize the conclusions of these papers to the case of nonlinear background perfect fluid with arbitrary $\omega_0\neq0$ and $\omega_1$. Now, let us investigate this general case in more detail. 

As it follows from Eqs.~\rf{3.9} and \rf{3.10}, in the limiting case ${r\to\infty}$, both $A^1$ and $B^1$ go to the Newtonian potential: $A^1(r\to\infty) = B^1(r\to\infty)=2\varphi_N/c^2=-2G_Nm/(c^2 r)$. The Newtonian gravitational constant $G_N$ is again connected with the multidimensional gravitational constant $\tilde{G}_{\mathcal{D}}$  by the formula \rf{2.21}, but this time for

\be{3.13}
\nu= -\frac{2\left[\Omega(1+\omega_0)-\omega_1\right]}{(1+\omega_0+2\omega_1)}\, .
\ee

\noindent
It can be easily seen that the relation $\nu=1$ is satisfied either for arbitrary $\Omega$ together with $\omega_0=-1$, or for $\Omega=-1/2$ and some arbitrary $\omega_0$. In the former case, our model reproduces asymptotic black branes presented in \cite{1202.2677}. Eqs.~\rf{3.11} and \rf{3.12} demonstrate that in the limit ${r\to\infty}$ the metric coefficient $G^1(r)$ vanishes asymptotically  to imply $\delta\varepsilon\rightarrow 0$ and the scalar curvature correction $R_1(r)$ becomes 
\be{3.14}
R_1(r)\to\frac{(2\Omega-1)}{c^2}\Delta_3\varphi_N \, .
\ee

\noindent
On the other hand, for the black brane value $\Omega=-1/2$ and arbitrary $\omega_0\neq 0$, owing to the identically vanishing Yukawa corrections, our solutions immediately read
\be{3.15}
A^1(r)=B^1(r)=2\frac{\varphi_N}{c^2} \, ,
\ee

\be{3.16}
R_1(r)=-\frac{2}{c^2}\Delta_3\varphi_N \, ,
\ee

\noindent
with $G^1(r)=0$.

It is worth noting that since the square root argument in Eqs.~\rf{3.9}-\rf{3.12} must be nonnegative, we need to impose the condition 

\be{3.17}
\frac{(1+\omega_0+2\omega_1)}{\omega_0}<0 \, ,
\ee

\noindent
which may be satisfied either for ${\omega_0<0}$ with ${1+\omega_0+2\omega_1>0}$, or for $\omega_0>0$ with $1+\omega_0+2\omega_1<0$. The former combination agrees with the above discussed case where we have set $\omega_0=-1$ and additionally, introduces the constraint $\omega_1>0$ which is the necessary condition for the internal space stabilization \cite{1107.3388,1202.2677,1402.1340}.  

Eq.~\rf{3.9} shows that in the case of arbitrary $\Omega\neq-1/2$, the gravitational potential acquires a Yukawa correction term. As the inverse square law experiments \cite{ISL} put restrictions on the parameters of such corrections, following the reasoning in \cite{1202.2677}, we may actually obtain an upper limit on the size of the extra dimensions here. For example, if $\omega_0=-1$ and for the natural assumption $|\Omega|$, $\omega_1\sim O(1)$, the maximal value of the internal space radius  $a_{\mathrm{max}} \sim 10^{-3}$cm \cite{ISL}. Given that the radius of the Sun $r_{\odot} \sim 7\times10^{10}$cm, we can drop the Yukawa correction terms in such models while studying the gravitational effects (the deflection of light and time delay of radar echoes)  with very high accuracy. 

\section{\label{Sec5}The case $\omega_0=0$}

We are now interested in the case $\omega_0=0$. Since $\omega_0\neq\bar\omega_0=-1$, the background perfect fluid is nonlinear.  According to Eq.~\rf{1.18}, a zero value of $\omega_0$ results in $\delta p_0=0$. The steps through \rf{1.19}-\rf{1.28} for this special case show that 
\be{4.1} 
0=-f'(R_0)\frac{G^1}{a^4}\,  ,
\ee

\noindent
which, for $f'(R_0)\neq 0$, implies
\be{4.2}
G^1=0\, .
\ee
Therefore, Eqs.~\rf{1.25}, \rf{1.26}, \rf{1.19}, \rf{1.21}, and \rf{1.28} read 
\be{4.3} \frac{1}{2} f'(R_0)\left(-A^1+B^1\right)-f''(R_0)R_1=0 \, ,\ee

\be{4.4}
\frac{f'(R_0)}{2}\left(\Delta_3B^1+R_1\right)+
\left(\Delta_3 R_1\right)f''(R_0) =0 \, ,
\ee
\ba{4.5}
\frac{f'(R_0)}{2}\left(\Delta_3A^1-R_1\right)-\left(\Delta_3 R_1\right)f''(R_0)=\kappa\left(
\delta\varepsilon +\hat\rho c^2\right) \, ,\nn\\
\ea
\ba{4.6}
f'(R_0)R_1&+&f''(R_0)\frac{2R_1}{a^2}+2 \left(\Delta_3 R_1\right)f''(R_0)\nn\\
&=&2\kappa\left(\delta p_1+\Omega\hat{\rho}c^2\right) \, ,\\ \nn
\ea
\ba{4.7}
-2f'(R_0)R_1+f''(R_0)\left(-\frac{2}{a^2} R_1-5\Delta_3R_1\right)\,\nn\\
=\kappa\left[\delta\varepsilon-2\delta p_1+\hat{\rho }c^2(1-2\Omega)\right] \, ,
\ea
respectively.

\subsection{$f^{\prime\prime}(R_0)\neq 0$}

We now demand that $f^{\prime\prime}(R_0)\neq 0$ and hence, exclude the linear model $f(R)=R+2\kappa\Lambda_6$. Combining Eqs.~\rf{4.6} and \rf{4.7}, it is possible to obtain an expression for $R_1$, that is

\be{4.8}
R_1=2\kappa a^2\left[\frac{\delta\varepsilon(1+3\omega_1)+\hat{\rho}c^2(1+3\Omega)}{6f''(R_0)+a^2f'(R_0)}\right] \, .
\ee

\noindent
Similarly, for $\Delta_3 R_1$, the same set yields

\ba{4.9}
\frac{\Delta_3 R_1}{\kappa}&=&-\frac{2f''(R_0)(\delta\varepsilon+\hat{\rho}c^2)}{f''(R_0)[6f''(R_0)+a^2f'(R_0)]} \, \nn\\
&&-\frac{a^2f'(R_0)\left[\delta\varepsilon(1+2\omega_1)+\hat{\rho}c^2(1+2\Omega)\right]}{f''(R_0)[6f''(R_0)+a^2f'(R_0)]} \, .
\ea

For further calculations, we assume a relation between $\delta\varepsilon$ and $\hat{\rho}c^2$ in the form
\be{4.10}
\delta\varepsilon=\zeta\hat{\rho}c^2 \, ,
\ee
which is the fine tuning condition. The parameter $\zeta$ here is to be specified in the later steps. Substituting  \rf{4.8} and \rf{4.9} into Eqs.~\rf{4.4} and \rf{4.5}, we obtain 

\ba{4.11}
&&\Delta_3 A^1(r)=2\kappa \hat{\rho}c^2 \,\nn\\
&&\times\left[\frac{4f''(R_0)(1+\zeta)+a^2f'(R_0)\left[1+\Omega+\zeta(1+\omega_1)\right]}{f'(R_0)\left[6f''(R_0)+a^2f'(R_0)\right]}\right] \, ,\nn\\
\ea
and
\ba{4.12}
\Delta_3 B^1(r)&=&2\kappa \hat{\rho}c^2 \,\nn\\
&&\times\left[\frac{2f''(R_0)(1+\zeta)-a^2f'(R_0)(\Omega+\zeta\omega_1)}{f'(R_0)[6f''(R_0)+a^2f'(R_0)]}\right] \, . \,\nn\\
\ea

\noindent
The overall factor $2\kappa\hat{\rho}c^2$ on the RHS of the above equations entails a delta-shaped function, therefore, one may immediately deduce that solutions $A^1(r)$ and $B^1(r)$ will both be proportional to $1/r$. Since we will eventually demand that $A^1(r\to\infty)=B^1(r\to\infty)$, it turns out useful to set \rf{4.11} and \rf{4.12} equal to each other from the very beginning. This requires 
\be{4.13}
\zeta=-\frac{2f''(R_0)+a^2f'(R_0)(1+2\Omega)}{2f''(R_0)+a^2f'(R_0)(1+2\omega_1)} \, ,
\ee

\noindent
hence, $R_1$ in \rf{4.8} reads

\be{4.14}
R_1=2\kappa a^2\left[\frac{\hat{\rho} c^2(\Omega-\omega_1)}{2f''(R_0)+a^2f'(R_0)(1+2\omega_1)}\right] \, .
\ee
In order to get the PPN parameter $\gamma=B^1/A^1=1$ as in General Relativity  (which is in very good agreement with the gravitational tests in the Solar system), from  Eq.~\rf{4.3}, we need to impose $R_1=0$. For vanishing $R_1$,  Eqs.~\rf{4.13} and \rf{4.14} imply
\be{4.15}
\Omega=\omega_1\, ,\quad \zeta =-1\,  .
\ee
In return, \rf{4.11} and \rf{4.12} take on the form
\be{4.16}
\Delta_3 A^1(r)=\Delta_3 B^1(r)=0 \, ,
\ee

\noindent
with the trivial solution 
\be{4.17}
A^1=B^1=0\, ,
\ee
taking into account the boundary condition
$A^1({r\to\infty})=B^1({r\to\infty})=0$. The physical reason for this trivial solution is that the matter sources $\delta\varepsilon$ and $\hat\rho c^2$ on the RHS of Eqs.~\rf{4.5}-\rf{4.7} mutually cancel: $\delta\varepsilon + \hat\rho c^2=0$.

\subsection{\label{Sec5.2}$f^{\prime\prime}(R_0)= 0$}

The linear model $f(R)=R+2\kappa\Lambda_6$ is a particular example of the case where $f^{\prime\prime}(R_0)=0$. Combined with the fine tuning condition \rf{4.10}, from \rf{4.6} and \rf{4.13}, we get
\be{4.18}
R_1=2\kappa\hat{\rho}c^2\left(\frac{\Omega+\zeta\omega_1}{f'(R_0)}\right) \, ,
\ee

\noindent
and
\be{4.19}
\zeta=-\frac{1+2\Omega}{1+2\omega_1} \, .
\ee
For this value of $\zeta$, Eq.~\rf{4.7} is reduced to Eq.~\rf{4.6}.
In the case $\Omega=0$, we obtain the relation $\zeta=-1/(1+2\omega_1)$, which exactly reproduces the corresponding results in paper \cite{1906.08214}. On the other hand, for black branes, i.e., for $\Omega=-1/2$, we find $\zeta=0$ that leads to $\delta\varepsilon=0$. For arbitrary $\Omega$, Eq.~\rf{4.3} demonstrates that the relation $A^1=B^1$ is satisfied automatically. As it follows from Eqs.~\rf{4.4} and \rf{4.5}, both of these corrections satisfy the equations
\be{4.20}
\Delta_3 A^1(r)=\Delta_3 B^1(r)=\kappa\hat{\rho}c^2\nu \, ,
\ee

\noindent
with solutions
\be{4.21}
A^1(r)=B^1(r)=-\frac{2}{c^2}\frac{1}{4\pi}\frac{S_D\tilde{G}_{\mathcal{D}}}{V_{\mathrm{int}}}\frac{m}{r}\nu \, ,
\ee
where
\be{4.22}
\nu\equiv\frac{2(\omega_1-\Omega)}{f'(R_0)(1+2\omega_1)}.
\ee

\noindent
The condition $A^1=B^1=2\varphi_N/c^2=-2G_Nm/(c^2 r)$ now imposes 
\be{4.23}
\frac{S_D\tilde{G}_{\mathcal{D}}}{V_{\mathrm{int}}}\nu=4\pi G_N \, .
\ee
For $f'(R_0)=1$, we see that black branes EoS $\Omega=-1/2$ restores the conventional relation $S_D\tilde{G}_{\mathcal{D}}/V_{\mathrm{int}}=4\pi G_N$. Obviously, the relation $A^1(r)=B^1(r)$ in \rf{4.21} guarantees that the PPN parameter $\gamma=B^1(r)/A^1(r)=1$ and exactly coincides with the value for $\gamma$ in General Relativity.
\section{\label{Sec6}Conclusion}

In the present paper we have studied multidimensional nonlinear $f(R)$ models with spherical compactification of the internal space. The background space-time is endowed with perfect fluid which makes the internal space compact and curved in the form of a two-dimensional sphere of radius $a$. Similar to the linear case \cite{1107.3388,1202.2677,1209.4501,1402.1340,ACZ}, this fluid has the vacuum-like EoS parameter $\bar\omega_0=-1$ in the external space. In the internal space,  the EoS parameter $\bar\omega_1$ depends on the form of $f(R)$. Then, we have perturbed this background by a massive gravitating source which is pressureless in the external space but has an arbitrary EoS parameter $\Omega$ in the internal space. We have considered a nonlinear perfect fluid, i.e., we have assumed that the parameters $\omega_0=\delta p_0/\delta \varepsilon\neq\bar\omega_0$ and $\omega_1=\delta p_1/\delta \varepsilon\neq\bar\omega_1$, where $\delta p_0$ ($\delta p_1$) denotes the fluctuation of the perfect fluid pressure in the external (internal) space and $\delta\varepsilon$ is the fluctuation in the corresponding energy density.

Then we have obtained the linearized Einstein equations for the perturbed metric coefficients. We have shown that all Einstein equations are reduced to the system of four master equations \rf{1.25} and \rf{1.31}-\rf{1.33}. Despite the rather complex form of these equations, we have found the exact solutions in the general case $f''(R_0)\neq0$, where $R_0$ is the scalar curvature of the background space-time and prime denotes differentiation with respect to the scalar curvature $R$. It was then revealed that the solutions are valid only for the case $\omega_0\neq0$. As is well known, a characteristic feature of nonlinear $f(R)$ models is the presence of the scalar degree of freedom, dubbed the scalaron. On the other hand, multidimensional models also have a scalar degree of freedom - the gravexciton/radion - due to the fluctuations of the internal space volume. Here, we have shown that the perturbed metric coefficients acquire correction terms in the form of the sum of two Yukawa potentials with Yukawa masses $\mu_1$ and $\mu_2$  \rf{2.18}.  We have demonstrated that masses $\mu_{1,2}$ have a rather complex form which cannot be related to the mass $m_{\mathrm{scal}}$ of scalaron and $m_{\mathrm{rad}}$ of gravexciton/radion by simple expressions. Nevertheless, for the concrete  model $f(R)=R+\xi R^2$, we were able to trace the relationship in a number of limiting cases. For example, when $|\xi|\sim a^2$, all masses appear to be of the same order of magnitude: $\mu_1,\mu_2,m_{\mathrm{scal}},m_{\mathrm{rad}}\sim 1/a$. A similar situation takes place for $|\xi|\gg a^2$. For $|\xi|\ll a^2$, we get $\mu_1\sim m_{\mathrm{rad}}$ and $\mu_2\sim m_{\mathrm{scal}}$. Another limiting case $a\to \infty$ corresponds to $\mu_1 = 0,\, \mu_2 = m_{\mathrm{scal}}$ and all the derived expressions for the metric coefficients exactly coincide with the formulas in paper \cite{1112.1539}, devoted to nonlinear models with toroidal compactification of the internal space. It has been demonstrated that the black string EoS $\Omega=-1/2$ ensures the fulfilment of the condition $A^1(r\to\infty)=B^1(r\to\infty)$.
	
In the degenerated case with a single Yukawa mass, $\mu_1=\mu_2=\mu$, metric coefficients no longer have the previously obtained form of combined Yukawa potentials. Instead, we have shown that the solutions have the form of a single Yukawa potential with a ``corrupted'' prefactor (in front of the exponential function) which, in addition to the standard $1/r$ term, 
contains a contribution independent of the three-dimensional distance $r$.

We have also investigated the class of models with $f''(R_0)=0$ (to which the specific linear model $f(R)=R+2\kappa\Lambda_6$ belongs) with nonlinear background perfect fluid for the condition $\omega_0\neq 0$. The linear gravity models with spherical compactification involving linear background perfect fluid were investigated previously in \cite{1107.3388,1202.2677,1209.4501}. In this work, we have generalized these models to the case of an arbitrary nonlinear background perfect fluid.
In agreement with the conclusions of previous papers, it has been shown that there are two possible situations in which the gravitational tests in the Solar system  (the deflection of light, time delay of radar echoes and perihelion shift) are satisfied and the corresponding constraints on the PPN parameter $\gamma$ are met. First, it takes place for a sufficiently large value of the gravexciton/radion mass $m_{\mathrm{rad}} \sim 1/a$, for which one may drop the Yukawa correction term in the metric coefficients. In this case, the EoS parameter $\Omega$ (that defines the pressure of the gravitating source in the internal space) can be arbitrary. Second, for the case $\Omega=-1/2$  \cite{1010.5740,1101.3910,1201.1756}, the Yukawa correction terms vanish automatically and arbitrary values of the mass $m_{\mathrm{rad}}$ are allowed. 

Finally, we have considered the case of zero speed of sound for the nonlinear perfect fluid in the external space: $\omega_0=0$. In the nonlinear gravity model with $f''(R_0)\neq0$, the condition  $\gamma=1$ for the PPN parameter $\gamma$ led to the trivial solution \rf{4.17} due to mutual cancellation of the matter sources on the RHS of the linearized Einstein equations \rf{4.5}-\rf{4.7}: $\delta\varepsilon + \hat\rho c^2=0$.  In the case $f''(R_0) = 0$, we have shown that the condition $\gamma=1$ is automatically  satisfied for arbitrary values of the EoS parameter $\Omega$. Yet, it is important to stress that only for the black branes/strings with $\Omega=-1/2$ one may ensure the absence of fluctuations in the background perfect fluid energy density, i.e. $\delta\varepsilon=0$, which indeed translates into the absence of nonphysical coat around the gravitating mass \cite{1202.4773}. 

The validity of our solutions is limited at two opposite scales. First, we consider the weak field limit in which the gravitational field is weak and peculiar velocities of test masses are much less than the speed of light. Consequently, the results cease to hold in the vicinity of such relativistic astrophysical objects as black holes and neutron stars.  Second, our approach is violated on scales where the cosmological expansion must be taken into account. The range of validity, in this case, depends on the masses of the astrophysical objects. For example, for a typical galaxy of a mass of the Milky Way, it corresponds to an order of 1 Mpc \cite{43,44}. In the four-dimensional case, an approach that allows to consider gravitational interaction on all scales (from astrophysical to cosmological ones) was proposed in paper \cite{45}.   

\appendix
  \renewcommand\thesection{Appendix}
\section{\label{appendix}Perturbed Ricci tensor}

In this appendix we present the nonzero components of the Ricci tensor corresponding to the metric \rf{1.11}. Obviously, the off-diagonal  components $R_{0i}$ $(i=1,2,3,4,5)$ vanish for static metrics. Using the results of the paper \cite{1107.3388}, we can easily find that  
the $R_{5i}$ components for $i=1,2,3,4$ and $R_{4i}$ components for
$i=1,2,3$ are also equal to zero. For the diagonal components we have

\ba{a.1}
R_{00} &= & \frac{1}{2}\Delta_3 A^1=R_{00}^{(1)}\, ,\nn\\
R_{ii} &= & \frac{1}{2}\Delta_3 B^1+\frac{1}{2}\left[-A^1+B^1+2\frac{G^1}{a^2}\right]_{x^ix^i}= R_{ii}^{(1)}\, ,\nn\\
i&=&1,2,3\, ,\nn\\
R_{44} &= & 1+\frac{1}{2}\Delta_3 G^1 = 1 + R_{44}^{(1)}\, ,\nn\\
R_{55} &= & \left(1+\frac{1}{2}\Delta_3 G^1\right)\sin^2\xi = \sin^2\xi + R_{55}^{(1)}\, , \ea

\noindent
where $\Delta_3=\partial^2/\partial x^2 + \partial^2/\partial y^2 +\partial^2/\partial z^2$ is the three-dimensional Laplace operator. The remaining off-diagonal components are
\ba{a.2}
R_{12} &= & \frac{1}{2}\left[-A^1+B^1+2\frac{G^1}{a^2}\right]_{xy}= R_{12}^{(1)}\, ,\nn\\
R_{13} &= & \frac{1}{2}\left[-A^1+B^1+2\frac{G^1}{a^2}\right]_{xz} =R_{13}^{(1)}\, ,\nn\\
R_{23} &= & \frac{1}{2}\left[-A^1+B^1+2\frac{G^1}{a^2}\right]_{yz} = R_{23}^{(1)}\, . \ea
The scalar curvature perturbation is a function of the external coordinates only: $R_1=R_1(x,y,z)$. The only non-zero derivatives are
\ba{a.3} (R_1)_{;i;k} \approx \frac{\partial^2 R_1}{\partial x^i\partial x^k} \quad \mbox{for}\quad i,k=1,2,3\, . \ea

\noindent
Then,
\be{a.4}
(R_1)_{;m;n}g^{(0)mn}=-\Delta_3 R_1\, .
\ee


\providecommand{\noopsort}[1]{}\providecommand{\singleletter}[1]{#1}%

\end{document}